# A new class of ferromagnetic semiconductors with high Curie temperatures

Nguyen Thanh Tu,[1,2] Pham Nam Hai,[1,3,4] Le Duc Anh,[1,5] and Masaaki Tanaka[1,4]

[1]Department of Electrical Engineering & Information Systems, The University of Tokyo, 7-3-1 Hongo, Bunkyo, Tokyo 113-8656, Japan.
[2]Department of Physics, Ho Chi Minh City University of Pedagogy, 280, An Duong Vuong Street, District 5, Ho Chi Minh City 748242, Vietnam
[3]Department of Electrical and Electronic Engineering, Tokyo Institute of Technology, 2-12-1 Ookayama, Meguro, Tokyo 152-0033, Japan.
[4]Center for Spintronics Research Network (CSRN), The University of Tokyo, 7-3-1 Hongo, Bunkyo, Tokyo 113-8656, Japan.
[5] Institute of Engineering Innovation, The University of Tokyo, 7-3-1 Hongo, Bunkyo, Tokyo 113-8656, Japan.

**Ferromagnetic semiconductors (FMSs), which have the properties and functionalities of both semiconductors and ferromagnets, provide fascinating opportunities for basic research in condensed matter physics and device applications. Over the past two decades, however, intensive studies on various FMS materials, inspired by the influential mean-field Zener (MFZ) model,[1,2] have failed to realise reliable FMSs that have a high Curie temperature ($T_C$ > 300 K), good compatibility with semiconductor electronics, and characteristics superior to those of their non-magnetic host semiconductors. Here, we demonstrate a new n-type Fe-doped narrow-gap III-V FMS, $(In_{1-x},Fe_x)Sb$, in which ferromagnetic order is induced by electron carriers, and its $T_C$ is unexpectedly high, reaching ~335 K at a modest Fe concentration $x$ of 16%. Furthermore, we show that by utilizing the large anomalous Hall effect of (In,Fe)Sb at room temperature, it is possible to obtain a Hall sensor with a very high sensitivity that surpasses that of the best commercially available InSb Hall sensor devices. Our results reveal a new design rule of FMSs that is not expected from the conventional MFZ model.**

Following the pioneering works on hole-induced ferromagnetism in Mn-doped III-V ferromagnetic semiconductors (FMSs), such as (In,Mn)As and (Ga,Mn)As,[3,4,5,6,7,8] intensive studies searching for reliable room-temperature FMSs have been conducted over the last 20 years. Many of the studies on FMSs have been inspired by the mean-field Zener (MFZ) model[1,2], which has been considered the "standard model" of FMSs. The most important prediction of this model was that wide-gap FMSs should have a much higher $T_C$ than narrow-gap FMSs. This prediction has been the "de facto" design rule for new FMS materials and led many researchers to focus on magnetically doped wide-gap semiconductors rather than narrow-gap ones. Despite worldwide efforts, however, there are three major problems that have not yet been solved. First, in the well-studied Mn-doped FMSs, there is no reliable n-type FMS available. Second, the highest Curie temperatures $T_C$ of (Ga,Mn)As (200 K) and (In,Mn)As (90 K) obtained so far are still much lower than room temperature, despite their very high hole concentrations ($p = 10^{20}$-$10^{21}$ cm$^{-3}$).[9,10] Third, the mechanism of their ferromagnetism and corresponding band structure are not completely understood and are still under debate.[1,2,11,12,13,14] Therefore, making FMSs that can be both n-type and p-type with room-temperature ferromagnetism remains a great challenge.

Recently, to overcome the shortcomings of the Mn-based FMSs, we have proposed and realized a new class of FMSs based on Fe-doped narrow-gap III-V semiconductors.[15-25] In Fe-doped III-V semiconductors, most Fe atoms replace the group-III sites and are in the isoelectronic $Fe^{3+}$ state. Thus, the Fe atoms play the role of local magnetic moments, while free carriers are supplied independently by co-doped non-magnetic donors/acceptors or native defects. This allows us to grow both n-type and p-type FMSs. Although the MFZ model predicts that it is impossible or difficult to prepare n-type FMSs due to their very weak s-d exchange interactions, we have demonstrated the first n-type FMS (In,Fe)As grown by low-temperature molecular beam epitaxy (LT-MBE).[15,16,17] In (In,Fe)As, we observed a $T_C$ of up to 70 K at an electron density ($n$) as low as $6 \times 10^{18}$ cm$^{-3}$. It was found that electron carriers reside in the conduction band with a small electron effective mass ($m^* = 0.03 \sim 0.17 m_0$, where $m_0$ is the free electron mass), high electron mobility (up to 1000 cmV$^{-1}$s$^{-1}$), and consequently long coherence length of the electron wavefunctions. In fact, we observed quantum size effects in (In,Fe)As quantum wells as thick as 40 nm and showed that the ferromagnetism in the quantum wells can be explained by the overlap of the electron wavefunctions and the Fe atoms.[18,19,20] We also demonstrated the "wavefunction engineering of ferromagnetism" in the (In,Fe)As quantum wells by changing the overlapping of the electron wavefunctions and the Fe atoms without large changes of the electron density.[20] Very recently, we observed large spontaneous spin-splitting (30 - 50 meV) in the conduction band of (In,Fe)As by using tunnelling spectroscopy in Esaki diode structures.[21] Furthermore, we realized a new p-type FMS (Ga,Fe)Sb [22,23] whose $T_C$ eventually reaches 340 K at $x = 25\%$.[24] Figure 1 summarises the highest $T_C$ values of Mn-doped (blue bars) and Fe-doped (red bars) III-V FMSs reported so far. The black circles and black diamonds show experimental $T_C$ values, and the black stars show the experimental $T_C$ values of the (In,Fe)Sb samples observed in this study. The $T_C$ values of Mn-doped III-V FMSs calculated by the MFZ model (white circles) with a Mn concentration of 5% and hole concentration of $3.5 \times 10^{20}$ cm$^{-3}$ are also shown. While the $T_C$ of Mn-based III-V FMSs (black circles) tends to decrease with the narrowing bandgap $E_g$ of the host semiconductor, the $T_C$ of Fe-based III-V FMSs (red diamonds) tends to increase, in contrast with the prediction of the MFZ model. We can therefore hypothesise that the ferromagnetic coupling in Fe-doped FMSs tends to increase with the narrowing bandgap. By extrapolating the trend of $T_C$ in several Fe-based FMSs, we anticipate that $T_C$ can reach room temperature most easily in (In,Fe)Sb, whose host InSb has the smallest bandgap of 0.17 eV among the well-established III-V semiconductors.

In this paper, to test our hypothesis and establish a new design rule for Fe-doped FMSs, we have grown and systematically investigated the crystal structure and magnetic, magneto-optical, and magneto-transport properties of $(In_{1-x},Fe_x)Sb$ thin films with various Fe concentrations ($x$ = 5 - 16%). We found that $(In_{1-x},Fe_x)Sb$ maintains the zinc-blende type crystal structure up to at least $x$ = 16%. The magnetic properties characterised by magnetic circular dichroism (MCD) spectroscopy and anomalous Hall effect (AHE) measurements confirmed the intrinsic ferromagnetism in (In,Fe)Sb. It is found that the $T_C$ of $(In_{1-x},Fe_x)Sb$ rapidly increases with $x$ and reaches ~335 K at $x$ = 16%. Furthermore, we show that the anomalous Hall effect of (In,Fe)Sb at room temperature can be used for Hall sensors, as its sensitivity surpasses that of the best commercially available InSb Hall devices.

**EXPERIMENTS**

**Growth and structural characterizations**

Figure 2(a) and Table I show the schematic structure and parameters of our samples, respectively. The $(In_{1-x},Fe_x)Sb$ layers were grown on semi-insulating GaAs(001) substrates by LT-MBE. A combination of 100-nm-thick AlSb / 10-nm-thick AlAs / 50-nm-thick GaAs buffer layers was used to reduce the lattice mismatch (11.1% – 11.6%) between the GaAs substrate and the $(In_{1-x},Fe_x)Sb$ layer (see Method for details of crystal growth). Samples A1-A5 with an $(In_{1-x},Fe_x)Sb$ thickness $d$ = 15 - 20 nm and Fe concentration $x = 5 – 16\%$ were prepared, as shown in the 1st - 3rd columns of Table I. The crystal structure of the (In,Fe)Sb layers was characterized by X-ray diffraction (XRD), scanning transmission electron microscopy (STEM), and transmission electron diffraction (TED). The XRD spectra show the zinc-blende crystal structure of (In,Fe)Sb with no other second phase such as FeSb and $FeSb_2$ precipitates, consistent with the *in situ* reflection high-energy electron diffraction (RHEED) patterns during the MBE growth (see Supplementary Information Section 1 for RHEED images and XRD spectra). Figure 2(b) shows the intrinsic lattice constant $a$ of $(In_{1-x},Fe_x)Sb$ as a function of $x$, estimated from the XRD spectra (see Supplementary Information Section 1 for the estimation of the lattice constant). One can see that $a$ linearly decreases with $x$, following Vegard's law. The best linear fitting of the data in Fig. 2(b) is given by the equation $a(x)$ [nm] $= 0.632(1-x) + 0.608x$. This means that the lattice constant $a_{FeSb}$ of the hypothetical zinc-blende FeSb crystal is 0.608 nm, which is close to that (0.592 nm) estimated from the lattice constants of (Ga,Fe)Sb[12] and (Al,Fe)Sb.[25] This indicates that most Fe atoms reside at the In site of the host InSb zinc-blende crystal. Figure 2(c) shows the STEM lattice image and transmission electron diffraction (TED) pattern (inset) of a representative sample A5 ($d = 15$ nm, $x = 16\%$) projected along the [110] axis. The STEM image and TED pattern clearly show that the crystal structure of the (In,Fe)Sb layers is of zinc-blende type without any visible second phase, which is consistent with the *in situ* RHEED and XRD observations. Figure 2(d) shows the corresponding energy dispersion X-ray (EDX) mapping for each element in this (In,Fe)Sb layer. We can see that there is no Fe segregation or precipitation and no void of In or Sb in the (In,Fe)Sb film, which would be observed if second-phase precipitates of Fe-Sb or Fe-In intermetallic compounds were formed. These results demonstrate that the epitaxial growth of zinc-blende (In$_{1-x}$,Fe$_x$)Sb is possible up to at least $x = 16\%$.

**Magneto-optical properties**

Next, we investigate the magneto-optical properties of the (In,Fe)Sb thin films using magnetic circular dichroism (MCD) spectroscopy in the reflection configuration. The MCD intensity in our study is expressed as $(90/\pi)[(R_+ - R_-)/(R_+ + R_-)] \propto \Delta E(1/R)(dR/dE)$, where $R$ is the reflectivity and $R_+$ and $R_-$ are the reflectivities for right ($\sigma^+$) and left ($\sigma^-$) circularly polarized light, respectively, $E$ is the photon energy, and $\Delta E$ is the Zeeman splitting energy. Since the MCD intensity is proportional to $dR/dE$ and $\Delta E$, it probes the spin-polarized band structure of the measured material. The MCD spectrum of an intrinsic FMS film would show the spectral features of the host semiconductor, with enhanced peaks at their optical critical point energies, whereas the MCD spectrum of a semiconductor film with embedded metallic magnetic precipitates would be broad without any particularly strong peaks related to the host semiconductor band structure. From the MCD spectral shape, we can judge whether the ferromagnetism comes from an intrinsic FMS or from magnetic precipitates. [26,27]

Figure 3(a) shows the MCD spectra of the (In$_{1-x}$,Fe$_x$)Sb films with $x = 5 - 16\%$, measured at 5 K with a magnetic field of 1 T applied perpendicular to the film plane. The MCD spectrum of a 20-nm-thick undoped nonmagnetic InSb film grown is also shown as a reference. The MCD spectrum of the InSb film shows very weak MCD signals, reflecting its small Zeeman splitting in InSb. In contrast, a large MCD (large Zeeman splitting) was observed in all the (In,Fe)Sb samples. All the MCD spectra of (In,Fe)Sb show two negative peaks at approximately 2.0 eV and 2.49 eV, corresponding to the critical point energies $E_1$ and $E_1 + \Delta_1$ of the InSb zinc-blende-type band structure, respectively.[28] Furthermore, there is no broad MCD background in any of the MCD spectra, confirming the absence of Fe-related metallic precipitates in our (In,Fe)Sb samples. These results agree well with the crystal structure analyses by XRD, STEM and TED shown in Fig. 2 and indicate that (In,Fe)Sb maintains the zinc-blende-type band structure of the host semiconductor InSb. Figures 3(b) shows the peak position of the $E_1$ optical transition energy, determined by fitting Lorentzian curves to the MCD spectra near $E_1$, as a function of $x$. Figure 3(c) shows the lattice constant $a$ as a function of the $E_1$ peak energy. One can see that the relation $E_1$ *vs.* $x$ follows Vegard's law, and the blueshift of $E_1$ is consistent with the decrease of $a$. This also confirms the successful growth of zinc-blende (In,Fe)Sb alloys. Figures 3(d) - 3(h) show the MCD intensity – magnetic field (MCD – $H$) curves of samples A1 – A5 ($x$ = 5 – 16%), measured at various temperatures 5 – 300 K. All the samples show open hysteresis, indicating the presence of ferromagnetic order. The saturation MCD intensity of (In,Fe)Sb systematically increases as $x$ increases. From the MCD – $H$ curves, we estimated the $T_C$ of (In,Fe)Sb using Arrott plots (see Supplementary Information Section 2 for details of the Arrott plots). The values of $T_C$ estimated by the Arrott plots are shown in Fig. 1 and listed in the 6th column of Table I. Figure 3(i) shows the $T_C$ of (In,Fe)Sb (red diamonds) and (Ga,Fe)Sb (blue dots,[22-24] for comparison) as a function of $x$. One can see that the $T_C$ of (In,Fe)Sb more rapidly increases as $x$ increases and reaches 335 K at $x$ = 16%. The $T_C$ values of (In,Fe)Sb are more than double those of (Ga,Fe)Sb at the same $x$. This confirms our hypothesis that the ferromagnetic coupling in Fe-doped FMSs increases upon narrowing the bandgap of the host semiconductor.

**Magneto-transport properties**

Next, we carried out magneto-transport measurements on the (In,Fe)Sb samples using patterned Hall bars (length: 200 μm, width: 50 μm). Figures 4(a) - 4(e) show the Hall resistance *vs*. perpendicular magnetic field ($R_{Hall} - H$) curves of samples A1–A5 ($x$ = 5 – 16%) at various temperatures. To eliminate the magnetoresistance contributions that are even functions of $H$, the odd-function contributions are extracted from the raw Hall data and plotted in (a) – (e). For samples A1–A4 ($x$ = 5 – 12%, $T_C$ = 60 – 250 K), the $R_{Hall} - H$ curves at low temperatures consist of a negative ordinary Hall resistance (OHR) and positive anomalous Hall resistance (AHR), but only the OHR was observed at room temperature. As determined from the sign of the OHE, samples A1 – A4 are n-type. For sample A5 ($x$ = 16%, $T_C$ ~ 335 K), however, the AHR is so large that the negative OHR cannot be seen even at room temperature. Note that the $R_{Hall} - H$ curves at low temperature of all the samples show very clear hysteresis characteristics, which are in good agreement with the MCD – $H$ characteristics shown in Figs. 3(d) - 3(h). The resistivity $\rho$ and electron concentration $n$, estimated by the Hall effect at 300 K, are listed in the 4th and 5th columns of Table I. Since samples A2– A5 ($x$ = 8 - 16%) show a high $T_C$ (150 – 335 K), we cannot estimate the electron concentration of these samples due to the large contribution of the AHR, even at 300 K. For sample A1 ($x$ = 5%, $T_C$ = 60 K), $n$ is estimated to be 3.2 × 10$^{17}$ cm$^{-3}$ at 300 K, which is 5 orders of magnitude lower than the doped Fe concentration. This result indicates that the Fe atoms in InSb are neither acceptors nor donors but remain isoelectronic, similar to the case of other Fe-doped FMSs (In,Fe)As and (Ga,Fe)Sb.

The observed ferromagnetism and large anomalous Hall effect at room temperature in sample A5 ($x$ = 16%, $T_C$ ~ 335 K) offer a new opportunity for application to magnetic field sensors. Under appropriate conditions, the anomalous Hall effect in the FMSs can be much stronger than the ordinary Hall effect, and thus it can be applied to Hall sensors. For realistic Hall sensors, two conditions must be satisfied. First, the FMS under consideration must be ferromagnetic at room temperature. Second, the bandgap of the FMS should be as small as possible to maximize the spin-orbit interaction. Neither condition can be satisfied in Mn-doped FMSs. In contrast, room-temperature ferromagnetism can be realized in narrow-gap p-type $(Ga_{1-x},Fe_x)Sb$ for $x \geq$ 23% and n-type $(In_{1-x},Fe_x)Sb$ for $x \geq 16\%$. In Fig. 4(f), we plot the voltage-related Hall sensor sensitivity $S$ of 250 Ω Hall sensors using (In,Fe)Sb ($x = 16\%$, red squares) and (Ga,Fe)Sb ($x = 25\%$, blue diamonds) as sensing materials, which are estimated from their anomalous Hall effect data at room temperature (See Supplementary Information Section 4 for the estimation of $S$). The inset in Fig. 4(f) shows a magnified view of the Hall resistance – magnetic field $H$ characteristics of sample A5. For comparison, we also plot the $S$ of three commercial Hall sensors made by Asahi Kasei Corporation using InSb (ultra-high sensitivity HW105-C, 250 Ω), InAs (high sensitivity HQ-0221, 370 Ω), and GaAs (typical sensitivity HG-0711, 650 Ω).[29] While the $S$ of (Ga,Fe)Sb is 0.25 mV.(mT.V)$^{-1}$ at 50 mT, the $S$ of our (In,Fe)Sb reaches 1.9 mV.(mT.V)$^{-1}$ at 50 mT, reflecting its strong AHE, probably caused by the strong spin-orbit interaction due to the smaller bandgap. Importantly, the $S$ of (In,Fe)Sb is higher than that of the commercial InSb Hall sensor with ultra-high sensitivity. This is the first demonstration of a characteristic of an FMS that is superior to that of its non-magnetic host semiconductor at room temperature. We note that the ordinary Hall effect is a classical electromagnetic phenomenon whose sensitivity is limited by the electron mobility of the sensing material, so there is little room for the improvement of the Hall sensor sensitivity. In contrast, the anomalous Hall effect in FMSs is a quantum mechanical phenomenon depending on the magnetization and the spin-orbit interaction of the FMS, which have much room for improvement by doping with more Fe or narrowing the bandgap. Our Fe-doped narrow-gap FMSs are very promising sensing materials for Hall sensors with ultra-high sensitivity.

**Electric-field control of ferromagnetism**

To further confirm the intrinsic ferromagnetism of (In,Fe)Sb, we demonstrate the electric-field control of the ferromagnetism in a field effect transistor (FET) structure of sample A4 ($x = 12\%$, $T_C = 250$ K). The FET device structure is illustrated in Fig. 5(a). Instead of a solid insulating layer, we use an electrolyte between the gate electrode and the (In,Fe)Sb channel of the FET, which forms an electric double layer (see Method for details of device fabrication). A fixed voltage $V_{DS}$ (= 0.5 V) was applied between the drain and source electrodes, and a gate voltage $V_{GS}$ was applied between the gate and source electrodes to modulate the channel electron density and the drain-source current $I_{DS}$. Figure 5(b) shows the $I_{DS} - V_{GS}$ curve measured at 220 K. The current $I_{DS}$ increases at a positive $V_{GS}$ and decreases at a negative $V_{GS}$, confirming that (In,Fe)Sb is $n$-type. Figure 5(c) shows the Hall resistance (HR) – $H$ characteristics measured at 220 K at $V_{GS}$ = -5, 0, +5 V. The hysteresis in the HR – $H$ characteristics is enhanced at $V_{GS}$= +5 and suppressed at $V_{GS}$ = -5 V. Figure 5(d) shows the change of the channel resistance $R_{xx}$ and the remanent Hall resistance $R_{(r)xy}$ as a function of $V_{GS}$. It is clear that the behaviour of the $V_{GS}$ dependence of $R_{(r)xy}$ is opposite that of $R_{xx}$. Since $R_{(r)xy} \sim (R_{xx})^{1.4}M_r$, where $M_r$ is the remanent magnetization (see Supplementary Information Section 5 for the derivation of this relationship), the $R_{(r)xy} - V_{GS}$ characteristic can only be explained by the $V_{GS}$ dependence of $M_r$. This can be seen more clearly in Fig. 5(e), where $M_r \sim R_{(r)xy}/(R_{xx})^{1.4}$ is plotted as a function of $V_{GS}$. These results demonstrate that the ferromagnetism of the (In,Fe)Sb thin film can be controlled by a gate electric field, confirming that the ferromagnetism is induced by electron carriers and that (In,Fe)Sb is an intrinsic FMS.

**DISCUSSION**

As an n-type FMS with the smallest bandgap of only 0.17 eV, the observation of ferromagnetism up to 335 K in (In,Fe)Sb at a modest Fe concentration of 16% is striking. This result firmly confirms that the $T_C$ of Fe-doped III-V FMSs tends to be higher in narrow-gap semiconductors, in contrast to the trend observed in many Mn-doped FMSs (see Fig. 1). Although the MFZ model of FMSs predicted that (1) it is very difficult or impossible to obtain *n*-type FMSs due to the very weak *s-d* exchange interactions and that (2) wide-gap FMSs should have much higher $T_C$ values than narrow-gap FMSs, our results overthrow these predictions, and a new theoretical model of ferromagnetism for FMSs that goes beyond the MFZ model is urgently required. We have also shown that the strong anomalous Hall effect in (In,Fe)Sb at room temperature can be used for Hall sensors whose sensitivity is superior to that of the commercially available InSb Hall devices. The applications of Fe-doped FMSs to realistic semiconductor spintronic devices at room temperature has become more realistic than ever.

# Method

**Crystal growth**

All of our samples were grown on semi-insulating GaAs(001) substrates by LT-MBE. First, we grew a 50-nm-thick GaAs buffer layer and a 10-nm-thick AlAs layer at a substrate temperature ($T_S$) of 550°C to obtain an atomically smooth surface. Next, we grew a 100-nm-thick AlSb buffer layer at $T_S$ = 470°C to reduce the lattice mismatch between (In,Fe)Sb and GaAs. Then, we grew an $(In_{1-x},Fe_x)$Sb layer with $x = 5 - 16\%$ at a growth rate of 0.5 µm/h at $T_S$ = 250°C. Here, the Fe flux was calibrated by secondary ion mass spectrometry and Rutherford back-scattering. The thickness of the $(In_{1-x},Fe_x)$Sb layer was fixed at $d = 20$ nm, except for that with $x = 16\%$, which was reduced to 15 nm to maintain good crystal quality and prevent phase separation. Finally, we grew a 2-nm-thick InSb cap layer to prevent the oxidation of the underlying (In,Fe)Sb layer. During the MBE growth, the crystallinity and surface morphology of the samples were monitored *in situ* by reflection high-energy electron diffraction (RHEED).

**Characterizations**

The high-resolution STEM analysis was performed by a JEM-ARM200F system (built by JEOL Corporation) with an acceleration voltage of 200 kV and an electron beam diameter of 0.1 nm. The STEM image was taken in high-angle annular dark field (HAADF) mode. The EDX element mapping was performed using a JED-2300T detector (also by JEOL). The MCD measurements were performed using a J700 system (built by JASCO Corporation) equipped with an electromagnet and a cryostat. AHE measurements were performed on samples etched into 50 × 200 µm Hall bars using photolithography and ion milling.

**FET device fabrication**

We fabricated electric double-layer-type field effect transistors (FETs) for the electrical control of ferromagnetism in (In,Fe)Sb. Sample A4 was patterned into a 50 × 200 μm Hall bar using photolithography and ion milling, and then a side-gate electrode and several terminal electrodes (source (S), drain (D), and electrodes for measuring Hall voltages) were formed by the electron-beam evaporation and lift-off of Au(50 nm)/Cr(5 nm) films. The side-gate electrode (G) and the (In,Fe)Sb channel were covered with the [N,N-diethyl-N-methyl-N-(2-methoxyethyl)ammonium bis(trifluoromethylsulfonyl) imide, DEME-TFSI] electrolyte to form a FET structure. As illustrated in Fig. 5(a), when a positive $V_{GS}$ is applied, the ions in the electrolyte accumulate at the surface of the semiconductor channel and form an electric-double-layer capacitor, which changes the electron density in the (In,Fe)Sb channel.

## Acknowledgements

This work is supported by Grants-in-Aid for Scientific Research (Grant No. 23000010, 16H02095, 15H03988) and the Yazaki Foundation. Part of this work was carried out under the Cooperative Research Project Program of RIEC, Tohoku University, and the Spintronics Research Network of Japan. N. T. T. acknowledges support from the JSPS Postdoctoral Fellowship Program (No. P15362). P. N. H. acknowledges support from the Toray Science Foundation.

## Author contributions

N. T. T. planned the research, grew the materials, characterized the MCD and AHE, fabricated the FET, and performed the electric-field control of the ferromagnetism. P. N.

H. planned and managed the research. N. T. T. and P. N. H. analysed the data. L. D. A. helped fabricate the FET. M. T. performed project planning and advised and managed the research. All authors wrote and commented on the paper.

**TABLE I. Structural parameters of our (In,Fe)Sb samples.** Thickness $d$, resistivity $\rho$, electron concentration $n$ at 300 K, and Curie temperature $T_C$ of $(In_{1-x},Fe_x)Sb$ samples A1 – A5 with various Fe concentrations $x = 5 - 16\%$ and layer thicknesses $d = 15 - 20$ nm.

| Sample | $x$ (%) | $d$ (nm) | $\rho$ ($\Omega$cm) | $n$ (cm$^{-3}$) | $T_C$ (K) |
|---|---|---|---|---|---|
| A1 | 5 | 20 | $2.6 \times 10^{-1}$ | $3.2 \times 10^{17}$ | 60 |
| A2 | 8 | 20 | $1.8 \times 10^{-1}$ | - | 150 |
| A3 | 11 | 20 | $1.4 \times 10^{-1}$ | - | 200 |
| A4 | 12 | 20 | $1.3 \times 10^{-1}$ | - | 250 |
| A5 | 16 | 15 | $1.6 \times 10^{-1}$ | - | ~335 |

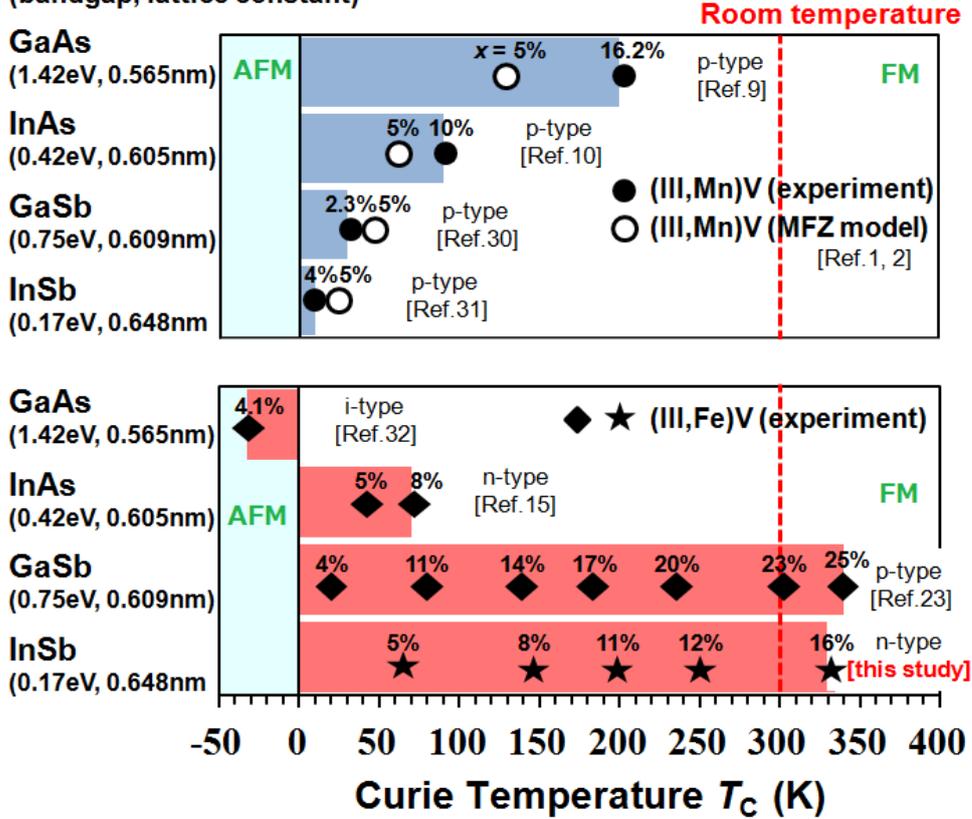

**Figure 1. Highest $T_C$ values of Mn-doped (blue bars) and Fe-doped (red bars) III-V FMSs reported so far.** Black circles and black diamonds show experimental $T_C$ values, and black stars shows experimental $T_C$ values of (In,Fe)Sb samples observed in this study. White circles show $T_C$ values of Mn-doped III-V FMSs calculated by the MFZ model with a Mn concentration of 5% and hole concentration of $3.5 \times 10^{20}$ cm$^{-3}$ (taken from refs. 1 and 2). Here, we plot the highest reported values of (Ga,Mn)As (ref. 9), (In,Mn)As (ref. 10), (Ga,Mn)Sb (ref. 30), (In,Mn)Sb (ref. 31), (Ga,Fe)As (ref. 32), (In,Fe)As (ref. 15), and (Ga,Fe)Sb (refs. 23, 24). The $T_C$ values of (In,Fe)Sb were obtained in the present work. AFM and FM are the antiferromagnetic and ferromagnetic coupling between the magnetic dopants, respectively. For (Ga,Fe)As, the paramagnetic Curie temperature estimated by the Curie-Weiss law is negative ($T_C < 0$), meaning that the exchange interaction between Fe atoms is antiferromagnetic (AFM).

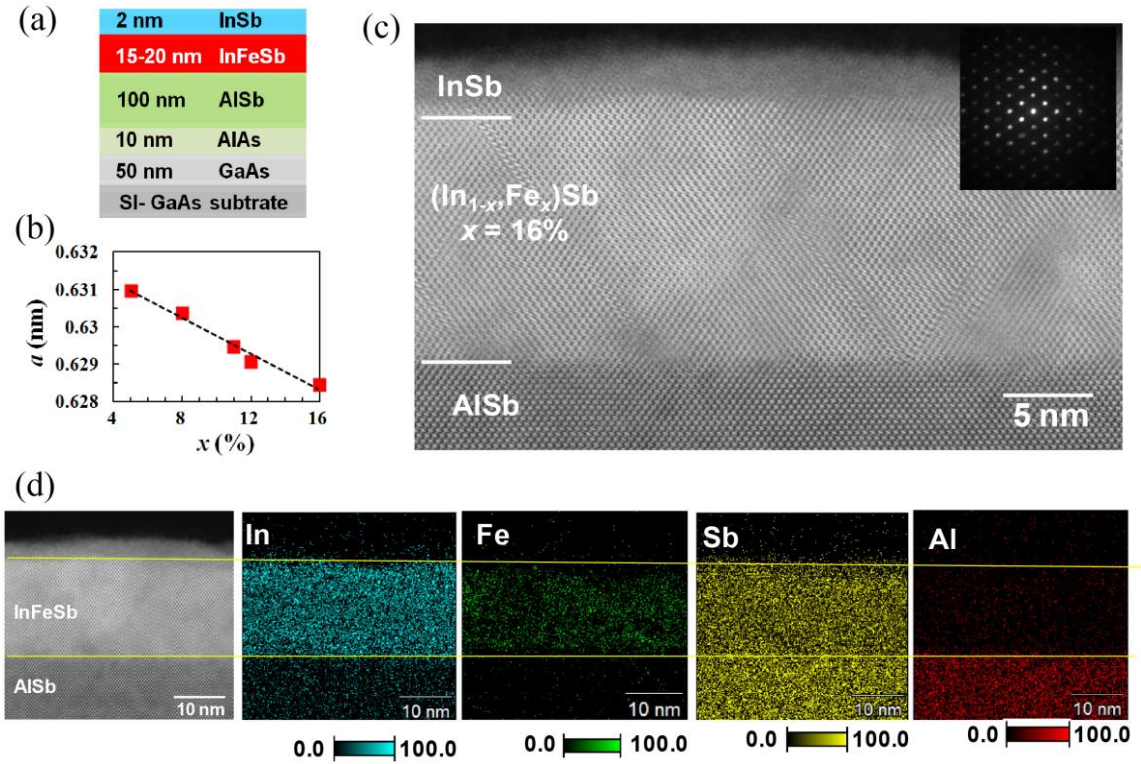

**Figure 2. Crystal structure analysis of (In$_{1-x}$,Fe$_x$)Sb samples A1 – A5 ($x$ = 5 – 16%).** (a) Schematic sample structure studied in this work. (b) Fe concentration ($x$) dependence of the lattice constant $a$ of (In$_{1-x}$,Fe$_x$)Sb measured by X-ray diffraction. (c) Cross-sectional scanning transmission electron microscopy (STEM) image of the 15 nm-thick (In$_{1-x}$,Fe$_x$)Sb layer in sample A5 ($x$ = 16%). Inset shows transmission electron diffraction (TED) of the (In$_{1-x}$,Fe$_x$)Sb layer. The STEM image and TED pattern indicate that the crystal structure of sample A5 ($x$ = 16%) is of zinc-blende type. (d) Energy dispersion X-ray (EDX) mapping for each element in the (In,Fe)Sb layer, which indicates that there is no precipitation or segregation of Fe atoms and no formation of Fe-In or Fe-Sb intermetallic compounds.

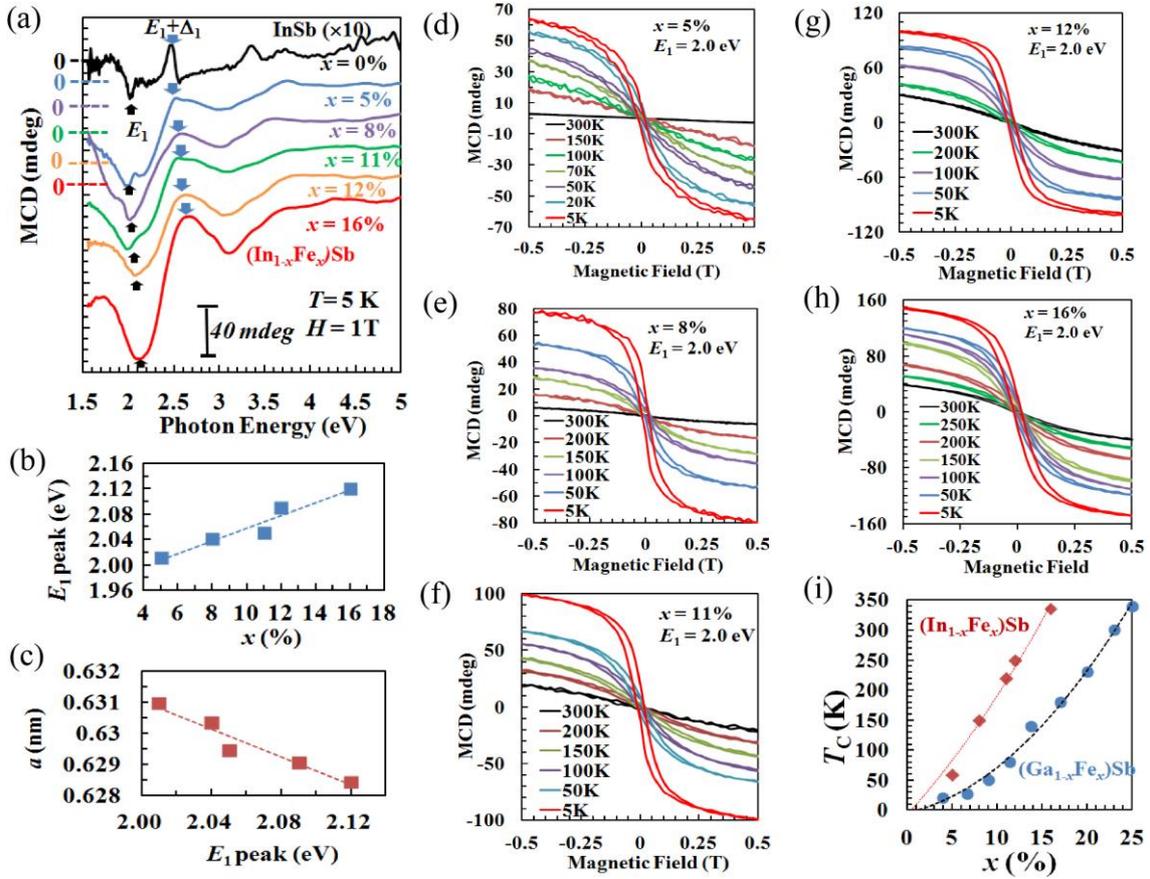

**Figure 3. Magneto-optical properties of (In,Fe)Sb.** (a) Reflection MCD spectra measured at 5 K under a magnetic field of 1 T applied perpendicular to the film plane for $(In_{1-x},Fe_x)Sb$ samples A1 – A5 ($x = 5 - 16\%$). The MCD spectrum of an undoped nonmagnetic InSb reference sample is also shown. (b) Fe concentration ($x$) dependence of the $E_1$ transition energy of the (In,Fe)Sb layers obtained from the MCD spectra. (c) Relationship between the lattice constant $a$ of the (In,Fe)Sb layers and their $E_1$ transition energy. (d) – (h) MCD – $H$ characteristics measured at a photon energy of 2.0 eV of the (In,Fe)Sb layers with a magnetic field applied perpendicular to the film plane. (i) $T_C$ of $(In_{1-x},Fe_x)Sb$ and $(Ga_{1-x},Fe_x)Sb$ as a function of $x$.

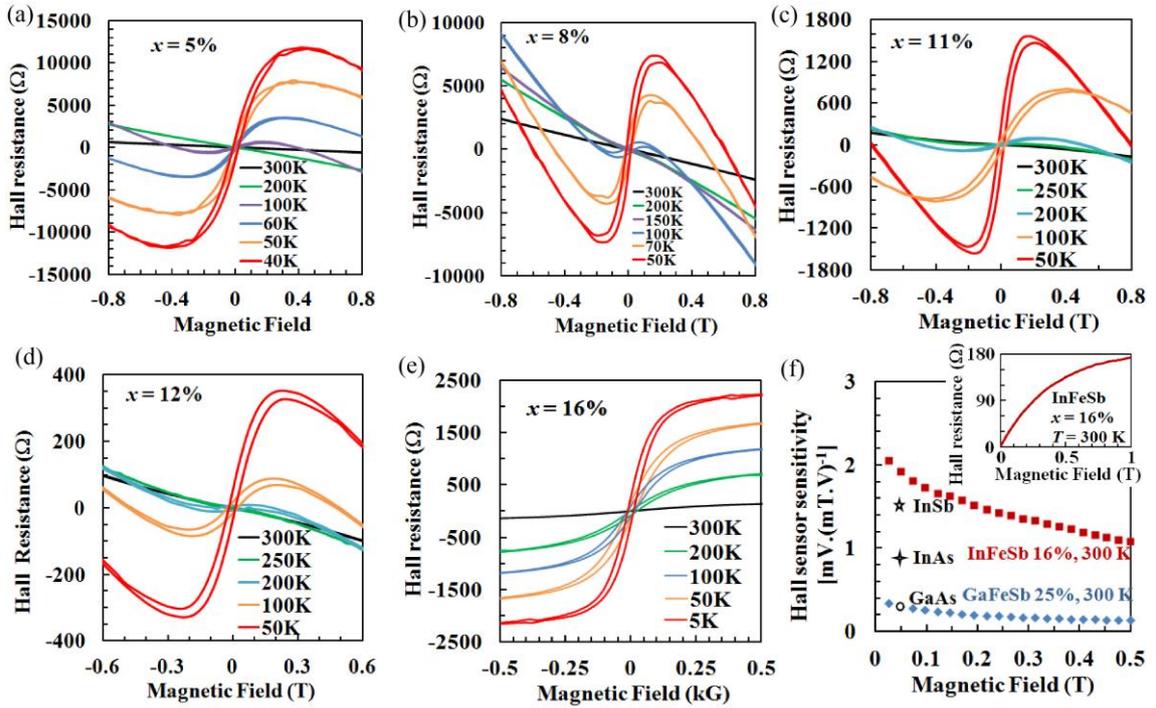

**Figure 4. Magneto-transport properties of (In,Fe)Sb**. (a) – (e) Hall resistance *vs.* magnetic field *H* applied perpendicular to the film plane of (In,Fe)Sb samples A1 – A5 ($x = 5 - 16\%$) measured at various temperatures. To eliminate the magnetoresistance contributions that are even functions of *H*, the odd-function contributions are extracted from the raw Hall data and plotted in (a) – (e). (f) Voltage-related Hall sensor sensitivity *S* of 250 Ω Hall sensors using (In,Fe)Sb ($x=16\%$, red squares) and (Ga,Fe)Sb ($x= 25\%$, blue diamonds) as sensing materials, estimated from their anomalous Hall effect data at room temperature. Inset shows a magnified view of the Hall resistance – *H* characteristics of sample A5 ($x = 16\%$). For comparison, we also plot the sensitivity *S* of three commercial Hall sensors from Asahi Kasei Corporation using InSb (ultra-high sensitivity HW105-C, 250 Ω), InAs (high sensitivity HQ-0221, 370 Ω), and GaAs (typical sensitivity HG-0711, 650 Ω).

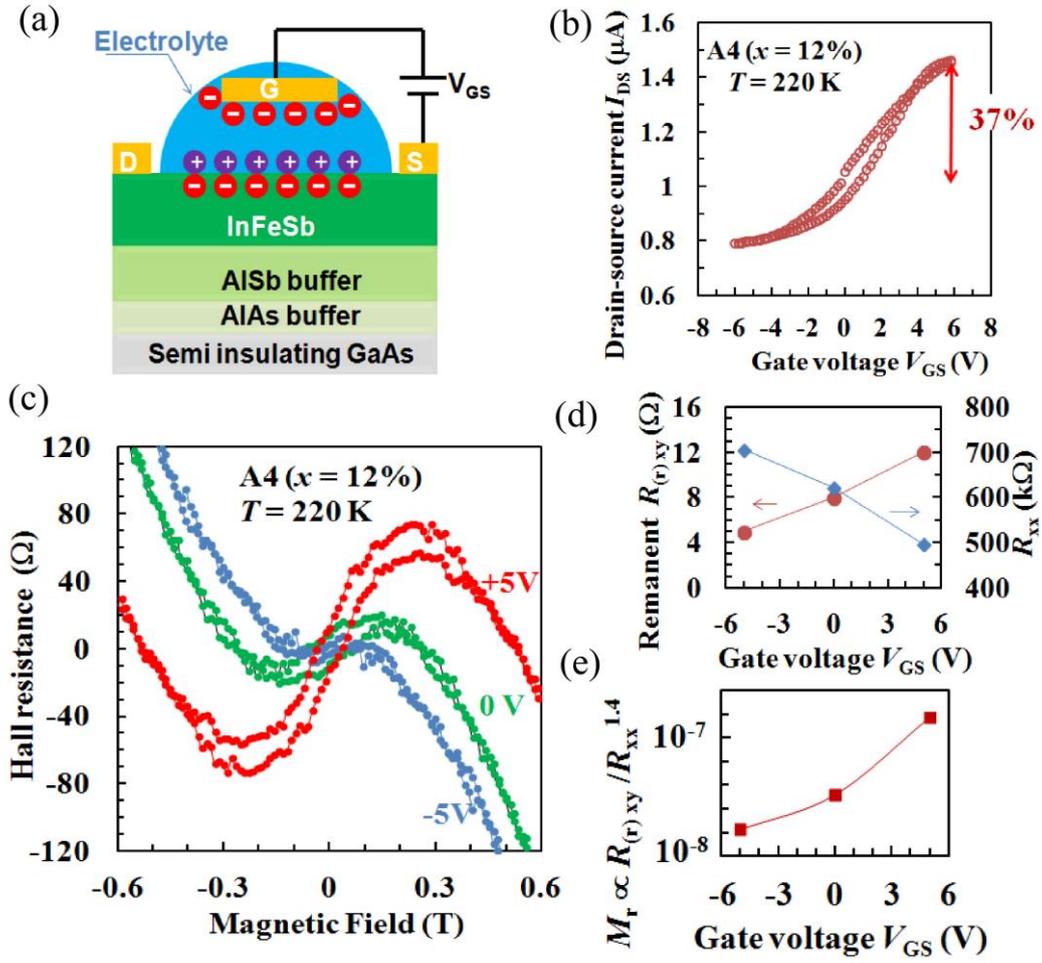

**Figure 5. Electric-field control of ferromagnetism in (In,Fe)Sb.** (a) Schematic structure of the FET device with electrolyte (DEME–TFSI) between the gate (G) and the (In,Fe)Sb channel. (b) Drain(D)–source(S) current ($I_{DS}$) – Gate voltage ($V_{GS}$) characteristics measured at 220 K for (In,Fe)Sb sample A4 ($x = 12\%$, $T_C = 250$ K). (c) Hall resistances measured at 220 K under various $V_{GS}$. (d) Remanent Hall resistance ($R_{(r)xy}$) and resistance ($R_{xx}$) measured at 220 K at $V_{GS} = +5, 0, -5$ V. (e) $V_{GS}$ dependence of $R_{(r)xy} / R_{xx}^{1.4}$, which is proportional to the remanent magnetization $M_r$.

# Supplementary Information

# A new class of ferromagnetic semiconductors with high Curie temperatures

Nguyen Thanh Tu,[1,2] Pham Nam Hai,[1,3,4] Le Duc Anh,[1,5] and Masaaki Tanaka[1,4]

[1]Department of Electrical Engineering & Information Systems, The University of Tokyo, 7-3-1 Hongo, Bunkyo, Tokyo 113-8656, Japan.
[2]Department of Physics, Ho Chi Minh City University of Pedagogy,
280, An Duong Vuong Street, District 5, Ho Chi Minh City 748242, Vietnam
[3]Department of Electrical and Electronic Engineering, Tokyo Institute of Technology
2-12-1 Ookayama, Meguro, Tokyo 152-0033, Japan.
[4] Center for Spintronics Research Network (CSRN), The University of Tokyo
7-3-1 Hongo, Bunkyo, Tokyo 113-8656, Japan.
[5] Institute of Engineering Innovation, The University of Tokyo
7-3-1 Hongo, Bunkyo, Tokyo 113-8656, Japan.

## 1. Reflection high energy electron diffraction (RHEED) patterns and estimation of the intrinsic lattice constant of (In,Fe)Sb from x-ray diffraction (XRD) spectra

Figures S1(b) – (f) show reflection high energy electron diffraction (RHEED) patterns taken along the $[\bar{1}10]$ azimuth of the $(In_{1-x},Fe_x)Sb$ layers of samples A1- A5 ($x$ = 5 – 16%) after the MBE growth of the (In,Fe)Sb layers. The RHEED patterns of (In,Fe)Sb are streaky 1 × 1 without any visible second phase, which are similar to the RHEED pattern of an undoped InSb sample as shown in Fig. S1(a). This suggests that (In,Fe)Sb layers grown by LT-MBE maintain the zinc-blende type crystal structure and have a smooth surface.

Figure S1(g) shows X-ray diffraction (XRD) of the $(In_{1-x},Fe_x)Sb$ samples A1 – A5 ($x$ = 5 – 16%). One can see only the diffraction peaks of the GaAs substrate, AlSb buffer and (In,Fe)Sb layers with zinc-blende crystal structures. No other phases such as FeSb and FeSb$_2$ were detected in the XRD spectra. This suggests that (In,Fe)Sb layers grown by LT-MBE maintain the zinc-blende crystal structure without any visible second phase, which is in good agreement with the RHEED patterns.

We estimate the intrinsic lattice constant of (In,Fe)Sb from the XRD spectra. The strained lattice constant along the *z*-axis [001] (growth direction) of the AlSb buffer ($a_{z,AlSb}$) and (In,Fe)Sb layer ($a_{z,(In,Fe)Sb}$) were estimated from the XRD spectra by using Gaussian fitting to their peak shapes. The strain of the AlSb buffer layers along the $z$-axis ($e_z$) was given by $e_z = (a_{z,AlSb} - a_{i,AlSb})/a_{i,AlSb}$, where the intrinsic lattice constant of cubic AlSb ($a_{i,AlSb}$) is 0.61355 nm.[1] From the obtained value of $e_z$, the in-plane strain of the AlSb layer ($e_x$) is estimated by $e_x = -2(C_{11,AlSb}/C_{12,AlSb})e_z$, where $C_{11,AlSb}$ and $C_{12,AlSb}$ are the second order elastic moduli of AlSb.[2] By using the value of $e_x$, the in-plane lattice constant of AlSb layers ($a_{x,AlSb}$) can be determined by $a_{x,AlSb} = (1+ e_x)a_{i,AlSb}$.

Next, we added two assumptions: (i) The (In,Fe)Sb layers are fully strained, meaning that the in-plane lattice constant $a_{x,(In,Fe)Sb}$ is equal to the in-plane lattice constant ($a_{x,AlSb}$) of the AlSb buffer layer. (ii) (In,Fe)Sb has the same elastic moduli as $C_{11,InSb}$ and $C_{12,InSb}$ of InSb.[3] Based on these assumption, the intrinsic lattice constant $a_{i,(In,Fe)Sb}$ of (In,Fe)Sb can be estimated by the following equation

$$a_{i,InFeSb} = [a_{z,InFeSb} + 2(C_{12,InSb}/C_{11,InSb})a_{x,InFeSb}]/[1+ 2(C_{12,InSb}/C_{11,InSb})] \quad (S1)$$

The intrinsic lattice constant $a_{i,(In,Fe)Sb}$ of $(In_{1-x}Fe_x)Sb$ obtained by Eq. (1) are plotted as a function of $x$ as in Fig. 2(b) in the main text.

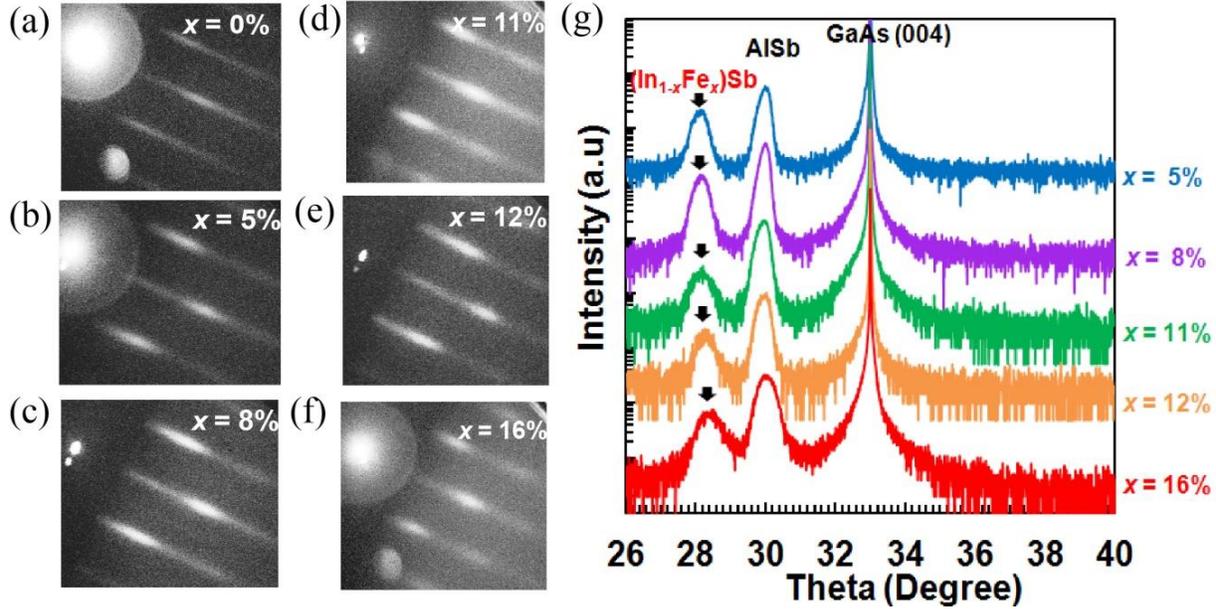

**Figure. S1.** (a) RHEED patterns taken along the $[\bar{1}10]$ azimuth after the MBE growth of the undoped InSb layers for a reference. (b) – (f) RHEED patterns taken along the $[\bar{1}10]$ azimuth after the MBE growth of the $(In_{1-x}Fe_x)Sb$ layers for samples A1 – A5 ($x$ = 5 – 16%, thickness $d$ = 15 – 20 nm). (g) X-ray diffraction (XRD) of $(In_{1-x}Fe_x)Sb$ samples A1 – A5 ($x$ = 5 – 16%) grown on GaAs (001) substrates.

## 2. Estimation of the Curie temperature $T_C$ of (In,Fe)Sb by using the Arrott plots of MCD – H characteristics

The $T_C$ values of all the (In,Fe)Sb samples were estimated by using the Arrott plots, which are $MCD^2 - H/MCD$ plots at different temperatures. Figures S2 (a) – (e) show the Arrott plots of the MCD – $H$ characteristics of 15 – 20 nm-thick (In,Fe)Sb samples A1 – A5 ($x$ = 5 – 16%). In Fig. S2(e), we cannot obtain the Arrott plots at $T$ > 320 K due to the limitation of our MCD measurement system. The $T_C$ value (335 K) of sample A5 ($x$ = 16%) were estimated by the extrapolation of the data measured at 300 – 320 K as shown in the inset of Fig. S2(e). The $T_C$ values are listed in the 4th column of Table I in the main text and plotted as a function of $x$ as shown in Fig 3(i) in the main text.

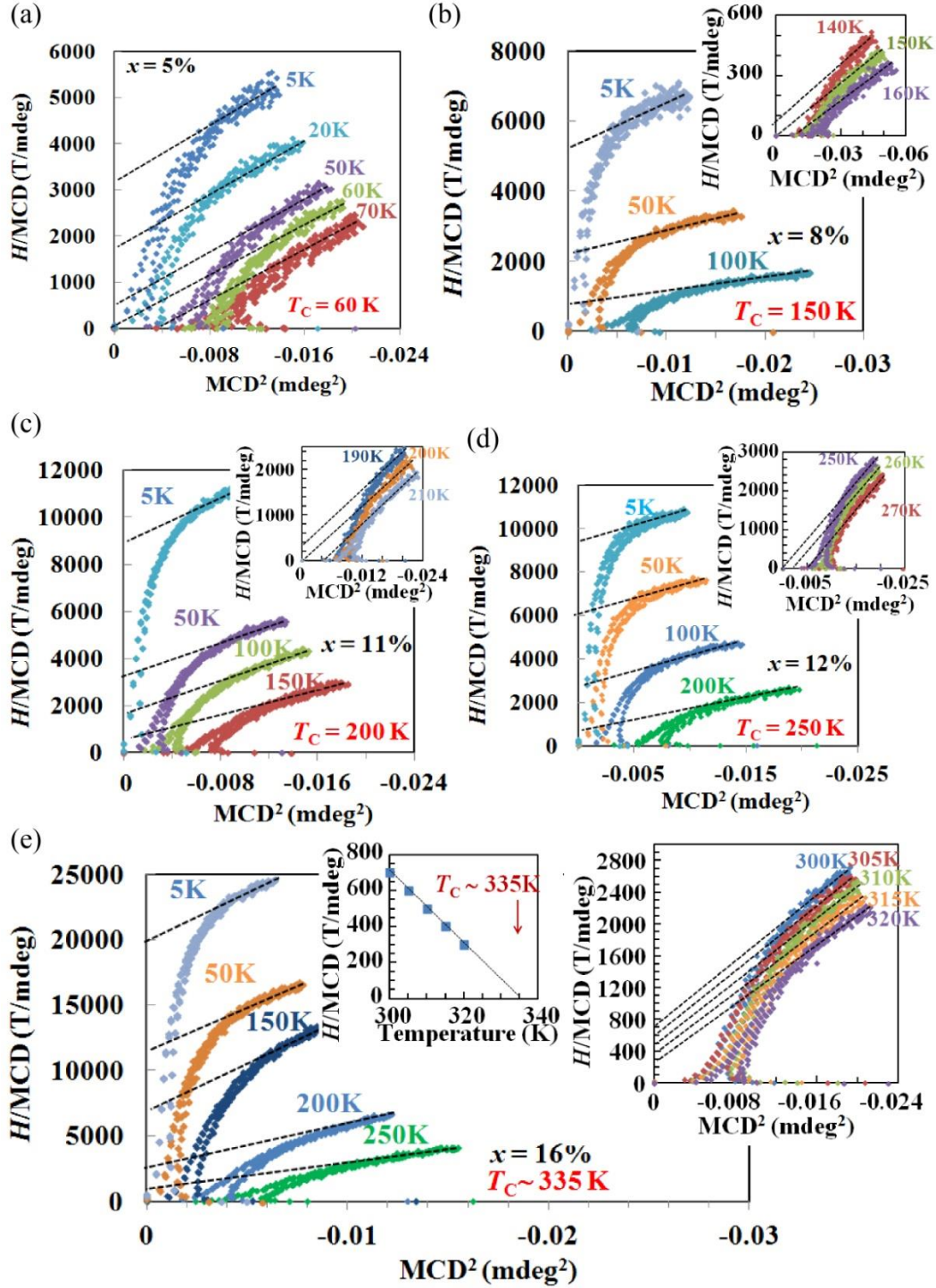

Figure S2. (a) – (e) Arrott plots of the MCD – $H$ characteristics of 15 – 20 nm-thick (In,Fe)Sb samples A1 – A5 ($x$ = 5 – 16%).

## 3. Normalized MCD spectra and MCD-$H$ characteristics

To study the magnetic properties of (In,Fe)Sb in more detail, we analyze the normalized MCD spectra and MCD – $H$ characteristics of three representative samples A1, A3, and A5 ($x$ = 5%, 11% and 16%). Figures S3 (a) – 3(c) show the MCD spectra normalized by their intensity at $E_1$ of these samples, measured at 5 K with various magnetic fields of 0.2, 0.5, and 1 T. The normalized MCD spectra of these samples show almost perfect overlapping on a single spectrum in the whole photon-energy range, indicating that the MCD spectra come from single-phase ferromagnetism of the entire (In,Fe)Sb film. We also see nearly perfect overlapping in their normalized MCD – $H$ characteristics measured at three photon energies [a ($E_1$ = 2.0 eV), b ($E_1$ + $\Delta_1$ = 2.39 eV), and c (3.0 eV)] shown in Figs. S3 (d) – (f), indicating again that the ferromagnetism in our (In,Fe)Sb comes from a single zinc-blende-type semiconductor phase, not from any second-phase precipitates.

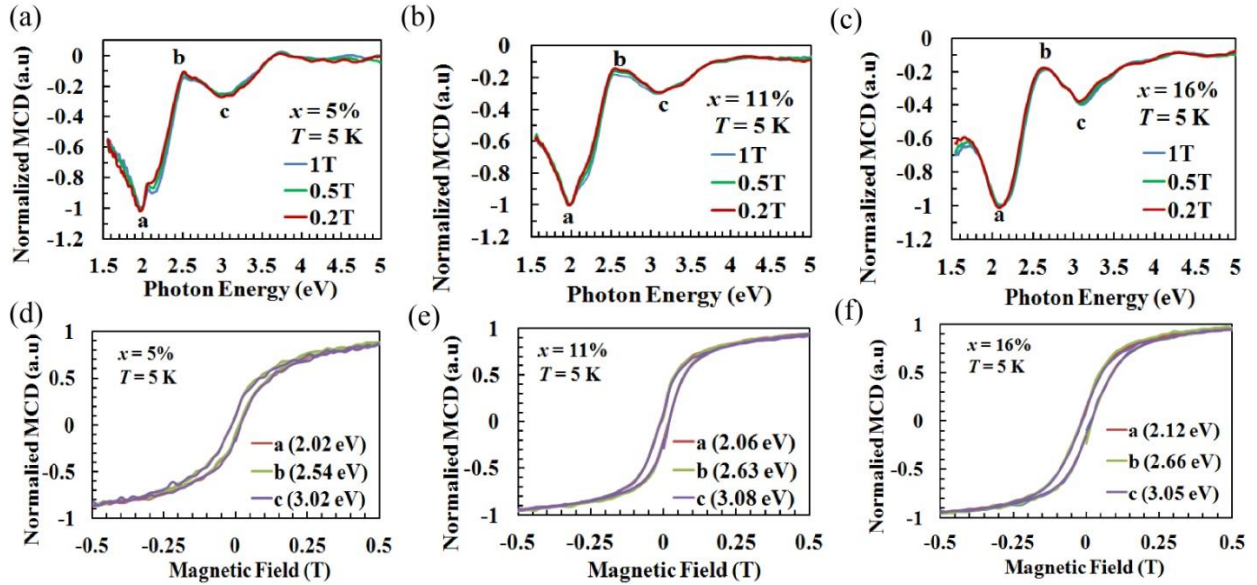

**Figure S3.** (a) - (c) Normalized MCD spectra of three representative ($In_{1-x}$,$Fe_x$)Sb samples A1, A3 and A5 ($x$ = 5%, 11%, and 16%) measured at 5 K under various magnetic fields of $H$ = 0.2, 0.5, and 1 T. Here, the MCD intensity at each photon energy is normalized by the value at $E_1$. (d) - (f) Normalized MCD–$H$ characteristics of samples A1, A3 and A5 ($x$ = 5%, 11%, and 16%) measured at various photon energies. Here, the MCD intensity at each magnetic field is normalized by the value at 1 T.

## 4. Estimation of the voltage-related Hall sensor sensitivity in the (In,Fe)Sb films

The Hall voltage $V_H$ of a Hall effect sensor using a thin semiconductor film is given by

$$V_H = \frac{1}{net} BI_B \qquad (S2),$$

where $e$ is the elementary charge, $n$ is the carrier density, $t$ is the thickness of the film, $B$ is the magnetic flux density, and $I_B$ is the bias current. For Hall sensors with high sensitivity, the sensing materials are usually n-type narrow-gap semiconductors with high electron mobility. Therefore, $n$ strongly depends on temperature. If the sensors are biased by a constant current source $I_B$, $V_H$ will show strong temperature dependence. To avoid this, Hall sensors with high sensitivity are usually biased by a constant voltage source $V_B$ rather than a constant current source. In this case, $V_H$ is given by

$$V_H = \mu \frac{W}{L} BV_B \qquad (S3),$$

where $\mu$, $W$, $L$ are the electron mobility, width, and length of the thin film, respectively. This equation shows that $V_H$ is less dependent on temperature when Hall sensors are biased by a constant voltage source. The most important figure of merit of voltage-biased Hall sensors is the voltage-related Hall sensor sensitivity $S$, defined as

$$S = \frac{V_H}{BV_B} \qquad (S4).$$

Note that, however, $S$ is not unique; it depends also on the sensor resistance. This is because lower sensor resistance allows more current at the same $V_B$, thus enhancing $V_H$ and $S$. Therefore, the sensor resistance must be mentioned when estimating $S$. Here, we show the $S$ values of three representative commercial Hall sensors from AsahiKasei corporation, whose datasheet can be found online.[4] The Hall sensor using InSb has the highest $S$ because of the high electron mobility of InSb compared with that of InAs and GaAs.

  InSb (ultra-high-sensitive HW105-C, 250 Ω): $S = 1.5$ mV.(mT.V)$^{-1}$

  InAs (high-sensitive HQ-0221, 370 Ω): $S = 0.87$ mV.(mT.V)$^{-1}$

  GaAs (typical-sensitive HG-0711, 650 Ω): $S = 0.25$ mV.(mT.V)$^{-1}$

Next, we estimate the voltage-related Hall sensor sensitivity $S$ of anomalous Hall effect sensors using (In,Fe)Sb or (Ga,Fe)Sb as sensing materials. Figure S4 shows the anomalous Hall voltage data at room temperature of a 200 μm × 50 μm Hall bar patterned from a 15 nm-thick (In,Fe)Sb thin film (Fe 16%, $T_C \sim 335$ K). The Hall bar resistance is 125 k$\Omega$. The Hall voltage was measured when the Hall bar was biased at $V_B = 6.25$ V and $I_B = 50$ µA, which is equivalent to that biased by $V_B = 1$ V and $I_B = 8$ µA. To directly compare with the InSb commercial Hall sensor (HW105-C, 250 $\Omega$, $V_B = 1$ V and $I_B = 4$ mA), we need to assume that the (In,Fe)Sb Hall sensor has the same resistance of 250 $\Omega$ and the same bias condition of $V_B = 1$ V and $I_B = 4$ mA. Such a Hall sensor can be realised by fabricating wider Hall bars with a more conductive (In,Fe)Sb layer. The Hall sensitivity of this 250 $\Omega$ (In,Fe)Sb Hall bar sensor can be estimated by multiplying that of the 125 k$\Omega$ Hall bar by a factor of 4 mA / 8 µA = 500. The results are shown in Fig. 4(f) in the main text.

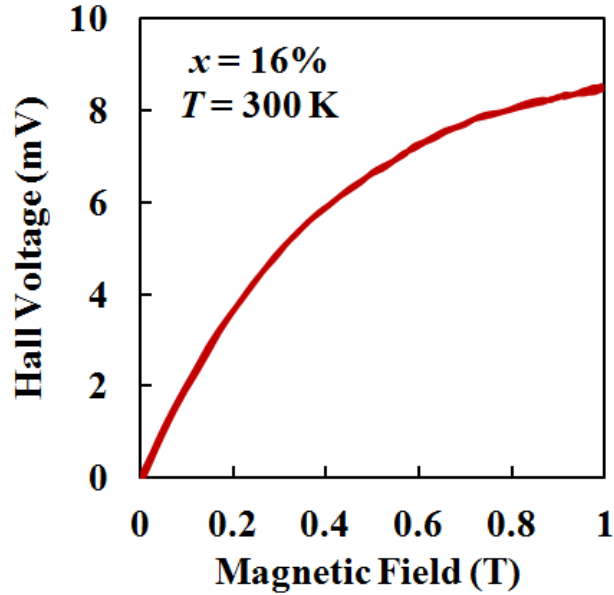

**Figure S4**. Anomalous Hall voltage data of a 15 nm-thick (In,Fe)Sb thin film (Fe 16%, $T_C \sim 335$ K), measured on a 200 µm × 50 µm Hall bar at room temperature. The device resistance is 125 k$\Omega$. The Hall voltage was measured when the device was biased by $V_B = 6.25$ V and $I_B = 50$ µA.

## 5. Relation of the anomalous Hall resistance and resistance

In order to study the relation of the anomalous Hall resistance and the longitudinal resistance in (In,Fe)Sb sample A4 ($x = 12\%$), we plotted the saturated value of the anomalous Hall resistance $R_{xy}$ of this sample as a function of the longitudinal resistance $R_{xx}$, measured at different temperatures (see Fig. S5). It is known that there is a relation $R_{xy} \propto R_{xx}^{\gamma}$, where $\gamma$ depends on the scattering mechanism.[5,6] The blue line in Fig. S5 shows the fitting $R_{xy} \propto R_{xx}^{1.42}$, which means $\gamma = 1.42$ for this sample.

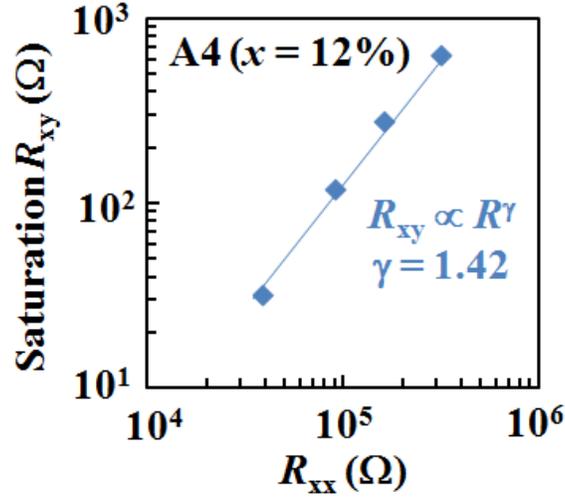

**Figure S5.** Relationship between the Hall resistance $R_{xy}$ and the longitudinal resistance $R_{xx}$ of the (In,Fe)Sb sample A4 ($x = 12\%$). The blue line shows the fitting line with $R_{xy} \propto R_{xx}^{1.42}$.

## 6. Magnetization

In order to further investigate the magnetic properties of (In,Fe)Sb, we measured magnetization by superconducting quantum interference device (SQUID) magnetometry. Two representative samples A3 ($x = 11\%$) and A5 ($x = 16\%$) were cooled from 300 K to 10 K with/without a magnetic field of 1 T perpendicular to the film plane [field cooling (FC)/zero field cooling (ZFC)]. Then, the magnetization was measured while increasing temperature with a small magnetic field of 30 Oe applied perpendicular to the film plane. Figures S6(a) and S6(b) show magnetization versus temperature ($M$ - $T$) curves of the two samples A3 ($x = 11\%$) and A5 ($x = 16\%$). For both field cooling (FC, blue) and zero field cooling (ZFC, red) conditions, $M$ - $T$ curves show similar monotonic behavior, which supports the single-phase ferromagnetism in these samples and is consistent with the normalized MCD spectra and normalized MCD – $H$ characteristics (see Supplementary Information Section 3). Due to the small remanent magnetization, a weak magnetic field of 30 Oe had to be applied during the $M$- $T$ measurement, which caused a tail up to temperature above $T_C$. Therefore, we estimate $T_C$ of these samples by using the Curie-Weiss plot at high temperatures. In Figs. S6(a) and S6(b), the green dots show the inverse of magnetic susceptibility $\chi^{-1} = H/M$ (at $H = 30$ Oe) at high temperatures, which follows the Curie-Weiss law. The $T_C$ values estimated from the Curie-Weiss plots are approximately ~220 K for sample A3 ($x = 11\%$) and ~335 K for sample A5 ($x = 16\%$), which are the same as those estimated by the Arrott plots of their MCD-$H$ characteristics (see Supplementary Information Section 2).

Figures S6(c) and S6(d) show the $M$-$H$ curves of samples A3 ($x = 11\%$) and A5 ($x = 16\%$) at 10 K, respectively, when the magnetic field was applied perpendicular to the film plane. The inset show the magnified figures near zero field. One can see that these $M$-$H$ curves show clear hysteresis, indicating the ferromagnetic order. The saturation magnetization and remanent magnetization increase with increasing $x$. The estimated saturation magnetic moment per Fe atom of these sample is 3.22 $\mu_B$/Fe atom for sample A3 ($x = 11\%$) and 3.39$\mu_B$/Fe atom for sample A5 ($x = 16\%$), which agree well with theoretical values calculated in Fe-doped InSb by first-principles electronic structure calculations (3.0 - 3.2 $\mu_B$),[7] and larger than those of Fe atoms (2.2 $\mu_B$) in Fe metal and Mn atoms (1.5 - 2.2 $\mu_B$) in (Ga,Mn)As.[8]

Figures S6(d) shows the $M - H$ characteristic for sample A5 ($x = 16\%$) at 300 K, which is close to its $T_C$ (~335 K). The $M - H$ characteristic show a hysteresis curve at 300 K, indicating the presence of ferromagnetic order at 300 K. The magnetization is significantly smaller than that at 10 K. This is consistent with $T_C$ ~ 335 K. If the ferromagnetism were governed by high-$T_C$ metallic phases such as Fe clusters, one would observe clear hysteresis with nearly the same magnitude of magnetization as those measured at 10 K. Therefore, it indicates that the ferromagnetism observed at room temperature in our (In,Fe)Sb samples is *not* due to such high-$T_C$ metallic precipitations.

For comparison, we show normalized $M - H$ and MCD $- H$ characteristics of the samples A3 ($x = 11\%$) and A5 ($x = 16\%$) at 10 K in the same figure, as shown in Fig. S6(e) and S6(f), respectively (the normalized $R_{Hall} - H$ curves are missing, since we cannot measure AHE at 10 K due to the high resistance of our samples at 10 K). The shape of the $M$-$H$ curve agrees well with the MCD $- H$ characteristic, indicating the single ferromagnetic phase in these samples. This result supports intrinsic ferromagnetism in our (In,Fe)Sb samples.

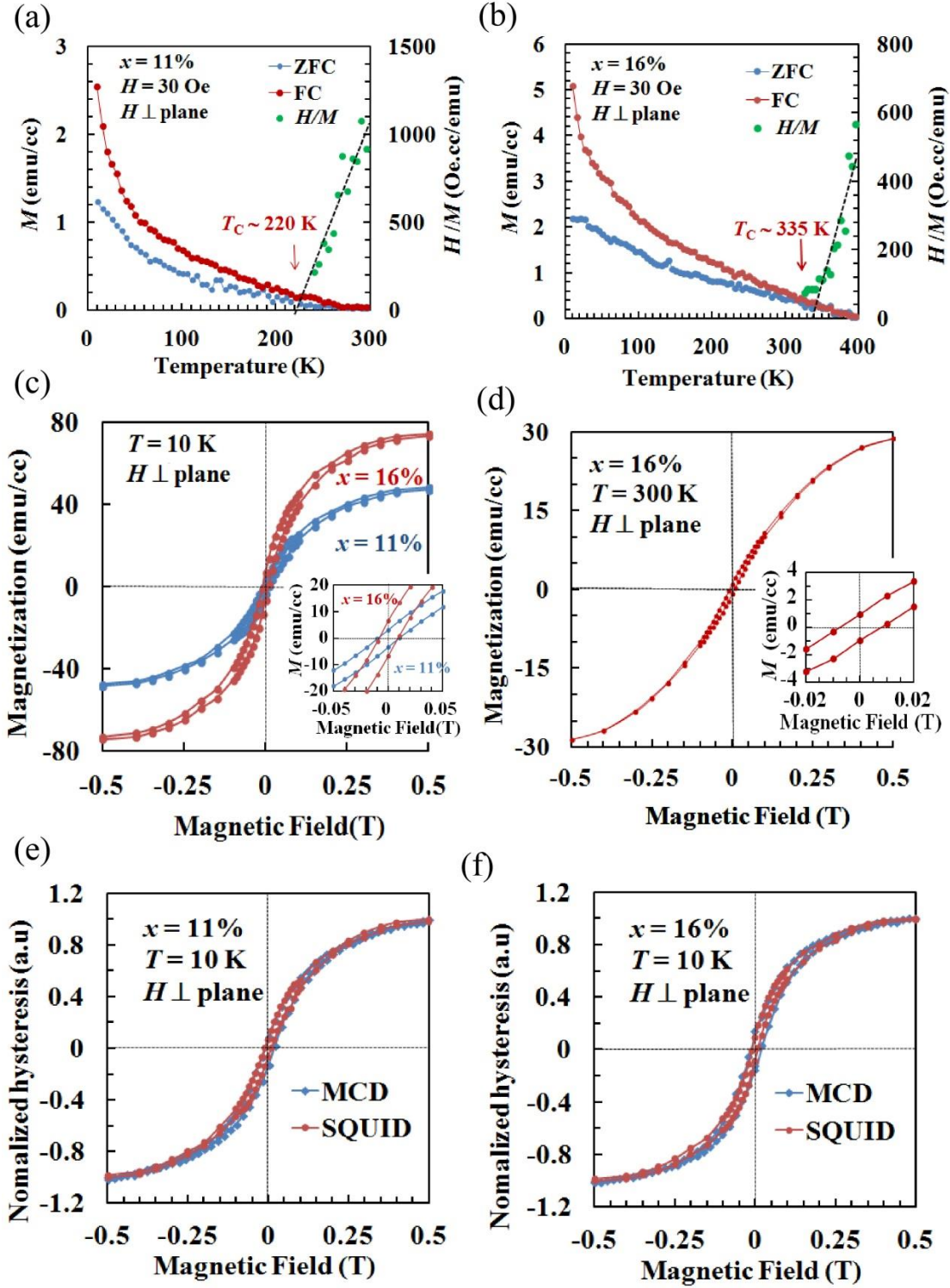

**Figure S6**. (a) and (b) Temperature dependence of the magnetization (*M - T* curves) of samples A3 (*x* = 11%) and A5 (*x* = 16%) measured by SQUID. The samples were cooled from 300 K to 10 K under a magnetic field of 1 T (FC, red dots) and under zero magnetic field (ZFC, blue dots). After cooling, the magnetization was measured with increasing temperature with a weak magnetic field of 30 Oe applied

perpendicular to the plane along the GaAs [001] direction.   (c) Magnetization hysteresis curves ($M – H$) of samples A3 ($x = 11\%$) and A5 ($x = 16\%$) measured at 10 K when the magnetic field was applied perpendicular to the plane. The inset in Fig. S6(c) shows the magnified view of the $M – H$ curves near zero field.   (d) Magnetization hysteresis curves ($M – H$) of sample A5 ($x = 16\%$) measured at 300 K when the magnetic field was applied perpendicular to the plane.   (e) and (f) Normalized $M – H$ and MCD – $H$ characteristics of samples A3 ($x = 11\%$) and A5 ($x = 16\%$) measured at 10 K, respectively.